\input{aipcheck}

\documentclass[
    ,final            
  ]
  {aipproc}
\usepackage{graphicx}
\usepackage{amssymb}

\layoutstyle{6x9}

\begin{document}

\title{Spin beyond Standard Model: Theory}

\author{J. Erler}
{
address={Departamento de F\'isica Te\'orica,
Instituto de F\'isica,
Universidad Aut\'onoma de M\'exico, \\
04510 M\'exico D.F.,
M\'exico}
}

\begin{abstract}
I use spin as a guide through the labyrinth of possibilities and ideas that go
beyond the established understanding of the fundamental interactions.
\end{abstract}

\maketitle


\section{Introduction}
What is required to define spin are Lorentz-covariant free particles and 
quantum mechanics. On the other hand, causality (from demanding a Lorentz 
invariant S-matrix) and the cluster decomposition principle (the requirement 
that distant experiments yield unrelated results which in relativistic theories
naturally leads to the concept of quantum fields) are not necessary. Thus, spin
can be considered even outside the framework of a quantum field theory (QFT) 
--- but not conversely --- and can serve as a very general organizing principle
for the many types of physics beyond the Standard Model (SM) that have been 
suggested\footnote{For tests of the SM see the contributions by Bill 
Marciano~\cite{Marciano} (muon anomalous magnetic moment and CKM unitarity), 
Yannis Semertzidis~\cite{Semertzidis} (electric dipole moments and 
$\mu \rightarrow e$~conversion), and Krishna Kumar~\cite{Kumar} (polarized 
electron scattering).}, including extra dimensions, strings, and M-theory.

The complete classification of unitary representations of the inhomogeneous 
Lorentz group according to Wigner~\cite{Wigner:1939cj} is recalled in 
Table~\ref{Wigner}.
\begin{table}[!b]
\begin{tabular}{lllll} 
\hline
  \tablehead{1}{r}{b}{$p^{2}$}          \hspace{16pt} &
  \tablehead{1}{c}{b}{$p_{0}$}          \hspace{16pt} & 
  \tablehead{1}{c}{b}{standard $k^\mu$} \hspace{16pt} &
  \tablehead{1}{c}{b}{little group}     \hspace{16pt} &
  \tablehead{1}{c}{b}{comment}          \hspace{16pt} \\
\hline
$> 0$ & $> 0$ & $(M,0,0,0)$ & $SO(3)$ & massive particle \\
$> 0$ & $< 0$ & $(-M,0,0,0)$ & $SO(3)$ & $E < 0$ (unphysical) \\
$= 0$ & $> 0$ & $(k,k,0,0)$ & $ISO(2)$ & massless particle \\
$= 0$ & $= 0$ & $(0,0,0,0)$ & $SO(3,1)$ & vacuum (no particles) \\
$= 0$ & $< 0$ & $(-k,k,0,0)$ & $ISO(2)$ & $E < 0$ (unphysical) \\
$< 0$ &   any  & $(0,M,0,0)$ & $SO(2,1)$ & tachyon (|v| > c)  \\
\hline
\end{tabular}
\caption{Wigner classification of unitary representations of the Poincar\'e 
group.}
\label{Wigner}
\end{table}
The little group is defined as the subgroup leaving some conveniently chosen 
"standard" four-momentum, $k^\mu$, unchanged and gives rise to spin and 
helicity. Thus, for a massive particle the possible spin states are obtained 
from the $SO(3)$ Lie algebra and are consequently identical to those familiar 
from non-relativistic quantum mechanics. The little group for massless 
particles is the Euclidean group in two dimensions, $ISO(2)$ (generated by two
translations and one rotation), and (being non-compact) permits, in general, 
a continuous parameter ("continuous spin"). One assumes (basically on 
phenomenological grounds) that physical states are non-trivially represented 
only with respect to the compact $SO(2) = U(1)$ subgroup, and identifies its 
"charges" with the helicity, $h$, of the massless particle. The trivial $U(1)$ 
Lie algebra would allow arbitrary values of $h$, but the topology of 
$SO(3,1) = SL(2,\mathbb{C})/Z_2$ can be shown to be that of the doubly 
connected space $\mathbb{R}^3 \times S^3/Z^2$, so that the requirement of 
a globally defined wave function restricts $h$ to integer and half-integer 
values. Still, Lorentz invariance requires consideration of the entire $ISO(2)$
little group. I will return to this point in the discussion of states with 
$h \geq 1$.

\begin{figure}[t]
\includegraphics[height=140pt]{mhmt_2007.eps}
\hspace{60pt}
\includegraphics[height=153pt]{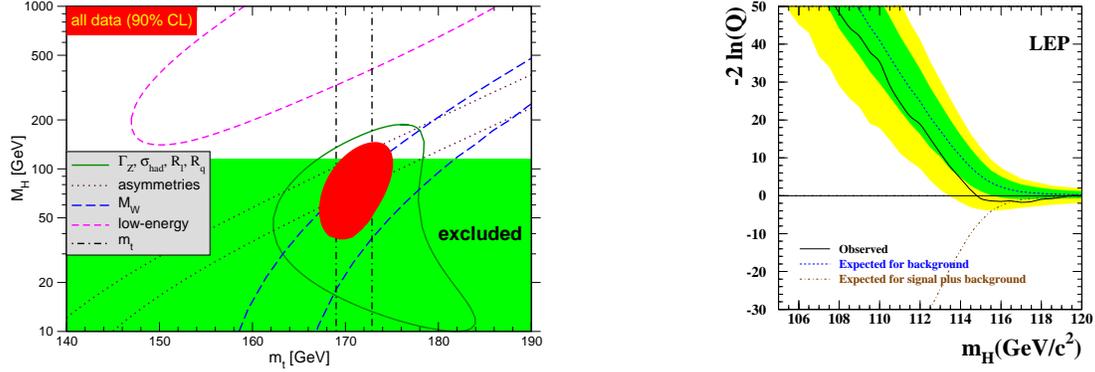}
\caption{Left: $1~\sigma$ contours from various data sets in the $M_H$-$m_t$ 
plane and 90\% region allowed by all precision data. Right: Results from Higgs 
searches at LEP~2~\cite{Barate:2003sz}. The bands around the dotted line for 
the background expectation refer to the 1~and 2~$\sigma$ regions. The solid 
line for the observation is below (above) the dotted line if an upward 
(downward) fluctuation is seen. A measure for the sensitivity to a given $M_H$ 
is the distance between the dot-dashed line for the signal plus background 
hypothesis and the dotted line.}
\label{mhmt}
\end{figure}

The spin~0 category includes the (would-be) Nambu-Goldstone bosons of 
spontaneously broken continuous symmetries, such as Higgs bosons, familons 
(broken family symmetry), Majorons (broken lepton number conservation), or 
axions (broken Peccei-Quinn symmetries). Furthermore, there may be radions 
(graviscalar components of the metric tensor in models with extra spatial
dimensions), dilatons ({\em e.g.\/}, in string theories), moduli (scalars with
flat potential in supersymmetry), inflatons, scalar leptoquarks, and sfermions 
(scalar superpartners). Spin~1/2 states beyond the SM could be due to a fourth 
family, right-handed neutrinos, other exotic states ({\em e.g.}, those needed 
to cancel anomalies in models with new gauge symmetries), techniquarks from 
dynamical (strong) symmetry breaking models, or X-inos (fermionic superpartners
other than gravitinos). Spin~1 states could be extra $Z^\prime$ or $W^\prime$ 
bosons or vector leptoquarks ({\em e.g.\/}, from Grand Unified Theories). 
Examples for spin~2 particles are the graviton ({\em predicted\/} by string and
M-theory) and its Kaluza-Klein excitations (in models with extra dimensions). 
A massless particle with spin $> 2$ is not expected to give rise to 
a long-range force, but the towers of massive string excitations include states
with arbitrarily high spins. On the other hand, a massless particle (or massive
particle after spontaneous supersymmetry breaking) of spin~3/2 would uniquely 
point to the gravitino of local supersymmetry (supergravity).

\section{Spin 0}
The unique fundamental scalar within the SM is the yet to be discovered Higgs 
boson. Its mass, $M_H$, is constrained from direct (so far negative) searches 
and from indirect precision analyzes (quantum loop effects). The global fit to 
all indirect data currently yields $M_H = 89^{+29}_{-22}$~GeV, with a very 
reasonable goodness of the fit. The $M_H$ constraints as functions of the top 
quark mass, $m_t$, are shown in Figure~\ref{mhmt} for various data sets. As can
be seen, these are consistent with each other with the exception of the low 
energy data which deviate mainly due to the result on deep inelastic neutrino 
scattering (NuTeV).

\begin{figure}[t]
\includegraphics[height=140pt]{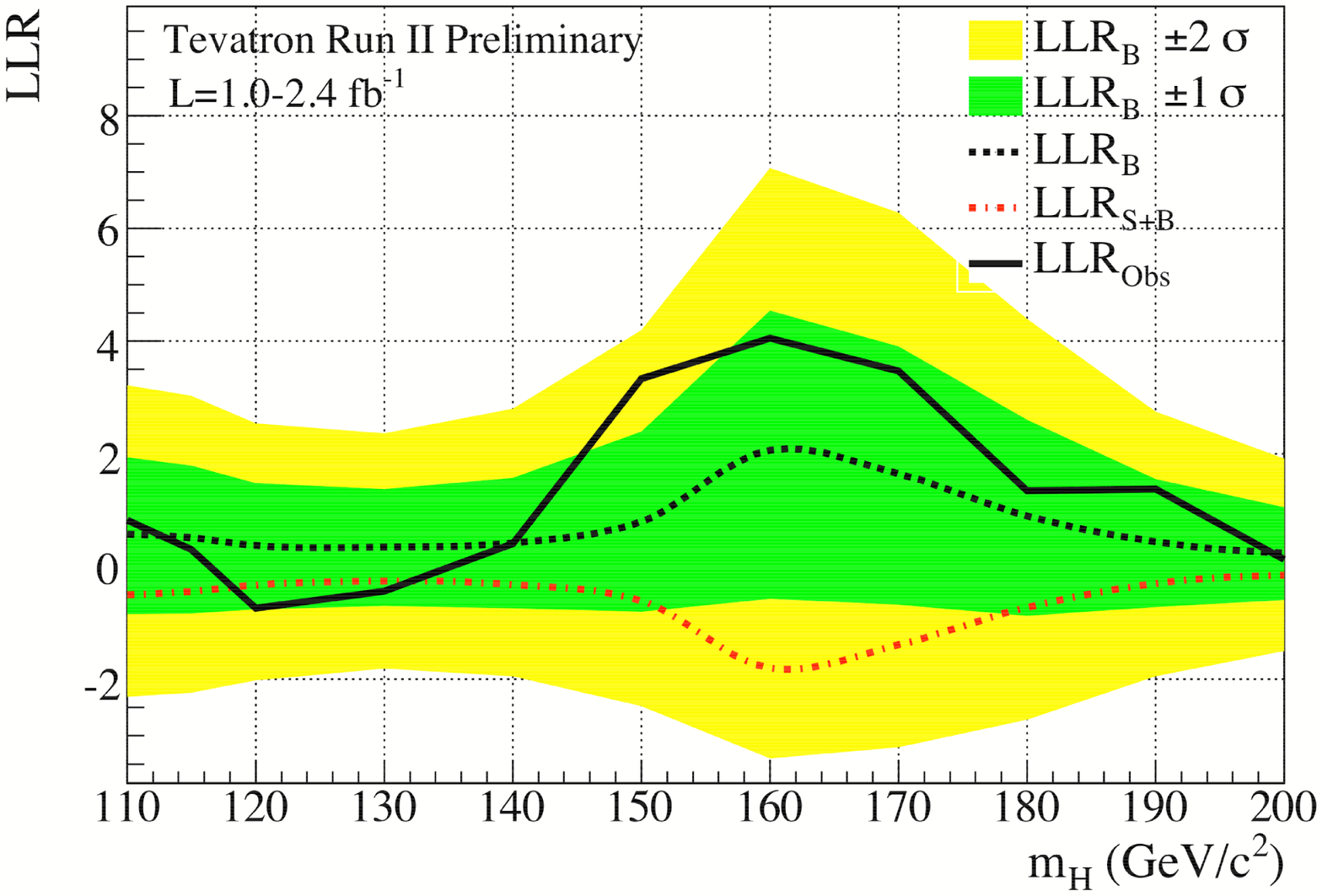}
\hspace{6pt}
\includegraphics[height=140pt]{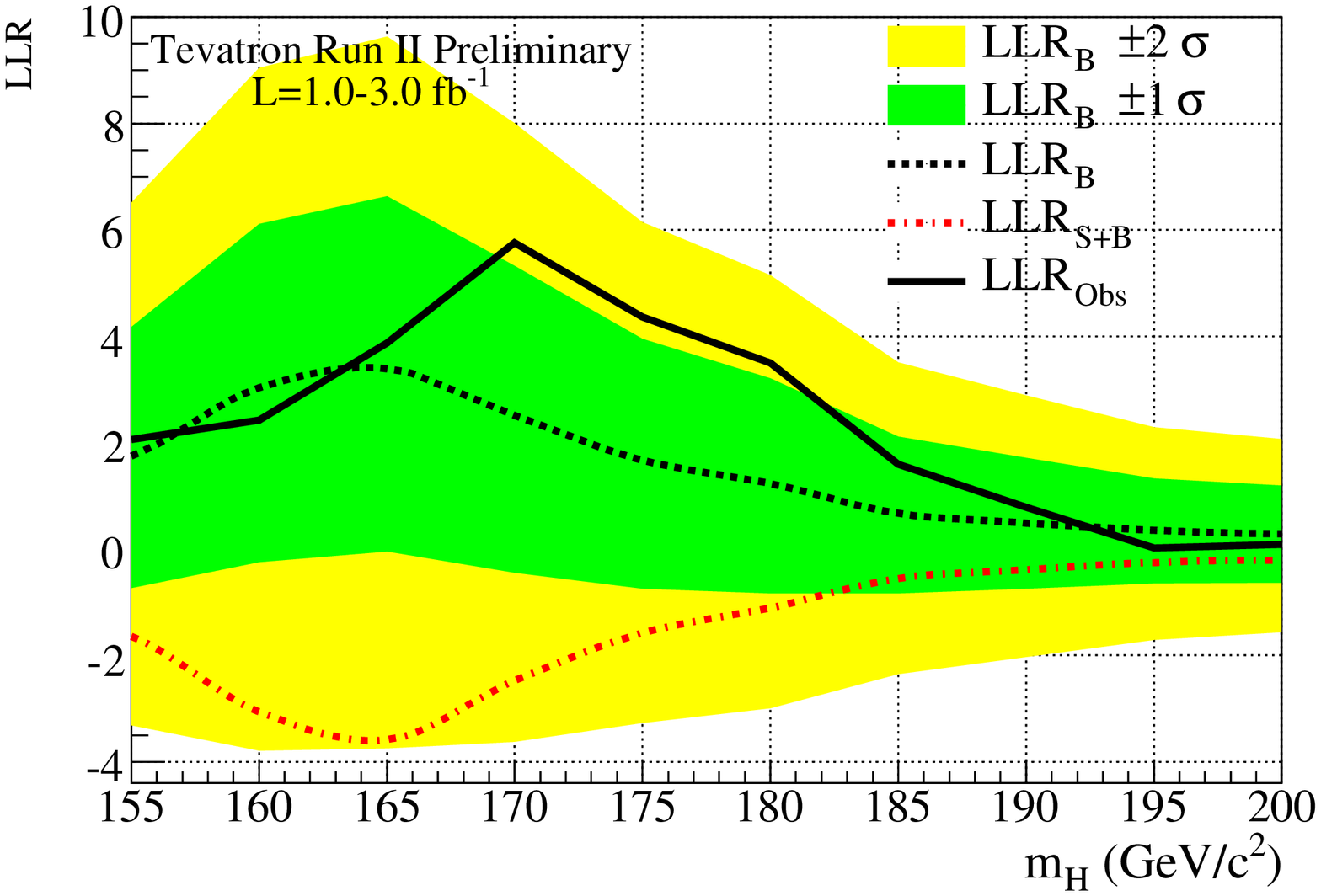}
\caption{Results from Higgs searches at run~II of 
the Tevatron~\cite{Bernardi:2008ee}. The meaning of the lines and bands is 
as in the right panel of Figure~\ref{mhmt}. Notice, that the right-hand plot is
from a slightly larger data set.}
\label{Tevatron}
\end{figure}

\begin{figure}[t]
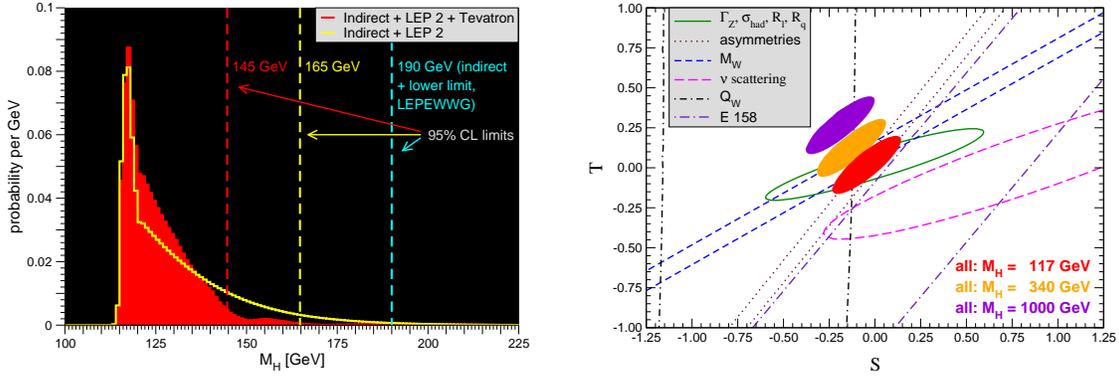

\includegraphics[height=140pt]{SPIN2008.eps}
\hspace{16pt}
\includegraphics[height=140pt]{ST_2007.eps}
\caption{Left: Combination of direct and indirect constraints on $M_H$. The red
(solid) histogram includes all data while the yellow line shows the result 
without the Tevatron constraints. The 95\% upper limits are also shown, where 
LEP~2 obtained their 190~GeV upper limit by treating their lower bound as 
a step function. Right: 1~$\sigma$ constraints on the parameters S and T 
(describing gauge boson self-energies) for various data sets and 
$M_H = 117$~GeV. The 90\% allowed regions for all data and three reference 
values for $M_H$ are also shown. In the SM, S = T = 0 (by definition) while in 
TC models typically S $\gtrsim 1$.}
\label{stplot}
\end{figure}

LEP~2 (Figure~\ref{mhmt}) excluded Higgs masses below 114.4~GeV (95\% CL), and 
saw a small but by itself insignificant excess around $M_H = 117$~GeV. 
An upward fluctuation is also seen for similar $M_H$ values at run~II of 
the Tevatron (Figure~\ref{Tevatron}). The best sensitivity is here for Higgs
masses slightly above the threshold for decays into $W$ pairs and thanks to 
a downward fluctuation values close to $M_H = 170$~GeV could already be ruled 
out. The combined result of all constraints (direct and indirect) is shown in 
Figure~\ref{stplot}. 

Even though the Higgs boson is a SM particle, it may provide clues about new
physics. {\em E.g.\/}, if one requires that it remains perturbatively coupled 
up to the fundamental Planck scale, $\kappa_4 \approx 2.4 \times 10^{18}$~GeV, 
one obtains (in the absence of new physics) the upper limit, 
$M_H \lesssim 180$~GeV. Likewise, vacuum stability up to $\kappa_4$ implies
$M_H \gtrsim 130$~GeV (if one allows a meta-stable vacuum this weakens to
$M_H \gtrsim 115$~GeV). In general, one expects $M_H$ to be of the form
$M_H^2 = ({M_H^0})^2+c{\alpha\over\pi}\kappa_4^2 \lll \kappa_4^2$, and
$c = {\cal O}(1)$ would require a delicate cancellation of the tree and
loop terms (the hierarchy problem\footnote{There is no hierarchy problem for 
fermions, because chiral symmetry protects massless fermions to acquire masses 
radiatively, and so the radiative corrections for massive fermions will be no 
larger than of the order of the chiral symmetry breaking tree mass terms 
themselves.}). The main mission of the LHC is then to find at least one Higgs 
boson and the new physics providing the solution to the hierarchy problem 
(ideally) also explaining (i) why the precision data appear to be consistent 
with the SM with so far no clear indication for new physics, (ii) the nature of
the dark matter, and (iii) the baryon asymmetry of the universe.

\section{Spin 1/2}
One way to address the hierarchy problem is to avoid fundamental scalars 
altogether. In the original {\em technicolor\/} (TC) 
idea~\cite{Weinberg:1979bn,Susskind:1978ms} (techni)fermions condense through 
non-Abelian gauge interactions, break electroweak symmetry (EWS), and generate 
the masses for the $W$ and $Z$ bosons. It is less straightforward to obtain 
the masses for the SM fermions and {\em extended technicolor\/} gauge 
interactions are needed. In general, this leads in turn to large flavor 
changing neutral current effects in conflict with observation. This can be 
cured by decelerating the running of the TC coupling (walking TC) so as to 
effectively decouple the extended TC gauge bosons. Another complication is 
the large $m_t$ value and one considers models of top quark condensation 
(topcolor). Models in which the EWS breaks solely due to topcolor predict too 
large an $m_t$ so that one arrives at hybrid models (topcolor assisted TC). One
also has to introduce an extra $U(1)^\prime$ gauge symmetry to prevent 
condensation of the much lighter bottom quarks. This is one of many examples in
which a model of EWS breaking predicts a specific massive $Z^\prime$ boson 
to solve some model building problem. Thus, $Z^\prime$ diagnostics may be 
an additional analyzing tool for the new physics that breaks EWS. However, 
models of TC are generally in conflict with the S parameter 
(Figure~\ref{stplot}). Incidentally, the S parameter also constrains a fourth 
fermion generation and rules it out if mass degenerate ($S = 2/3\pi$). 
The non-degenerate case is also disfavored and only marginally consistent for 
specific mass values.

Another way to avoid fundamental scalars is to assume that the Higgs field is 
{\em composite\/}~\cite{Kaplan:1983sm}. A modern reincarnation of this idea is
{\em Little Higgs Theory\/}~\cite{ArkaniHamed:2001nc} with a fundamental scale
$\Lambda$ of 5 to 10~TeV, and where the Higgs is lighter since it appears as a
pseudo-Goldstone boson. In these models the quadratically divergent 
contributions to $M_H$ are then postponed by one loop order (in some models by
two orders). One can also assume that quarks and leptons are composite. This 
introduces effective contact interactions with an effective scale constrained 
by LEP~2 to satisfy $\Lambda \gtrsim {\cal O}(10\mbox{ TeV})$.

\section{Spin 1}
In a QFT, a particle with $h = \pm 1$ can produce a long-range ($1/r^2$) force 
only if it is coupled to a conserved vector current. This can be traced to 
the ISO(2) little group mentioned in the introduction and implies the concept 
of gauge invariance~\cite{Weinberg:1965rz}. Extra neutral gauge bosons 
($Z^\prime$) can arise from extra $U(1)^\prime$ gauge symmetries as contained, 
{\em e.g.}, in the $E_6$ unification group, 
$E_6 \to SO(10)\times U(1) \to SU(5)\times U(1)^2$, or in left-right symmetric 
models, $SU(2)_L\times SU(2)_R\times U(1) \to SU(2)_L\times U(1)^2$. 
The $Z^\prime$ could also be a Kaluza-Klein (KK) excitation (sequential 
$Z^\prime$), the techni-$\rho$, or the topcolor-$Z^\prime$ mentioned above. 
The $Z^\prime$ mass, $M_{Z^\prime}$, is constrained to be greater than 1305~GeV
in the sequential case (DELPHI) and $M_{Z^\prime} > 630$ to 891~GeV for $E_6$ 
scenarios (D\O), while electroweak (EW) data limit the mixing angle with 
the ordinary $Z$ to $< 0.01$. Likewise, mass and mixing of an extra $W^\prime$ 
are limited, respectively, to $> 1$~TeV (D\O) and $< 0.12$ (OPAL).

\section{Spin 2}
In a QFT, $1/r^2$-forces are produced by an $h = \pm 2$ particle only if it is
coupled to the conserved energy-momentum tensor implying invariance under 
general coordinate transformations~\cite{Weinberg:1965rz}. The ultraviolet
completion of gravity requires its embedding into structures like 
{\em string\/} or {\em M-theory\/} whose growing number of known vacua led to 
the concept of the {\em string landscape\/}~\cite{Susskind:2003kw} and 
a revival of anthropic reasoning (the multiverse). 

String~\cite{Green:1987sp} and M-theory~\cite{Horava:1995qa} predict 
the existence of extra dimensions 
(EDs), some of which are conceivably much larger than $\kappa_4^{-1}$. One can 
consider~\cite{ArkaniHamed:1998rs} a factorized 
$D$-dimensional space, $\mathbb{R}^4 \times M^{D-4}$, where $M$ is 
a {\em flat\/} space with volume $(2\pi R)^{D-4}$. The $D$-dimensional Planck 
scale, $\kappa_D$, is identified with the EW scale, and the hierarchy problem 
appears here as the puzzle why 
$R^{-1} = \kappa_D (\kappa_D/\kappa_4)^{2/(D-4)} \ll \kappa_D$ (the case 
$D = 4 + 1$ is excluded from precision studies of Newton's law). Collider 
processes such as $e^+ e^- \to \gamma + $~missing energy require 
$\kappa_D > {\cal O}(\mbox{1~TeV})$, while astrophysical bounds,
$\kappa_D \gtrsim {\cal O}(100\mbox{ TeV})$, can be avoided in models.
An alternative is to introduce a {\em warped\/} 5-dimensional
space~\cite{Randall:1999ee} with metric, 
${\rm d} s^2 = e^{-yk} \eta_{\mu\nu} {\rm d} x^\mu {\rm d} x^\nu - {\rm d}y^2$,
where $k$ is the anti-de-Sitter (AdS) curvature and $0 \leq y \leq \pi R$ 
the fifth coordinate. The EW scale is here exponentially suppressed
($\sim \kappa_4 e^{-\pi k R}$). Observables like the $\ell^+ \ell^-$ or 
$\gamma\gamma$ invariant mass spectra can probe the interaction scale of KK
gravitons with matter concluding $k R \gtrsim 11$.

In addition to the graviton one can also allow SM fields to propagate in 
the factorized EDs. {\em E.g.\/}, TeV-scale string 
compactification~\cite{Antoniadis:1990ew} with compactification radius
$R \gtrsim (7 \mbox{ TeV})^{-1}$ (from LEP~2) implies gauge fields propagate in
the bulk. {\em Gauge-Higgs Unification\/}~\cite{Manton:1979kb} protects the EW 
scale by defining the Higgs as a higher-dimensional gauge field component.
{\em Universal Extra Dimensions\/} (UEDs)~\cite{Appelquist:2000nn} allow all SM
particles in the bulk. Grand Unified Theories (GUTs) in EDs with
$R^{-1} \approx M_U$ (the gauge coupling unification scale) can solve many 
problems of conventional supersymmetric GUTs~\cite{Hall:2002ci}. There are also
models with SM fields propagating in warped EDs (but not the Higgs in order 
to protect the exponential suppression of the EW scale). These can be viewed as
{\em dual\/} descriptions of walking TC by virtue of the AdS correspondence
with conformal field theories~\cite{Maldacena:1997re}. No fully realistic 
models exist, but it is interesting that EWS breaking by boundary conditions of
gauge fields in warped EDs yield {\em higgsless models\/}~\cite{Csaki:2003dt}.

\section{Spin 3/2}
In a QFT, an $h = \pm 3/2$ particle can produce a $1/r^2$-force only if it is 
coupled supersymmetrically. Conversely, supersymmetry~\cite{Wess:1973kz} is 
the only possibility to extend the Poincar\'e algebra non-trivially and its 
non-renormalization theorems offer the most elegant solution to the hierarchy 
problem. Moreover, in models of supersymmetric unification a large $m_t$ 
correctly predict EWS breaking (a fact that was 
known~\cite{AlvarezGaume:1981wy,Ibanez:1982fr} even before $m_t$ was measured) 
and these are further supported by the approximate unification of gauge 
couplings at $M_U \sim 10^{-16} \mbox{ GeV} \lesssim \kappa_4$ in the minimal
supersymmetric standard model (MSSM). Finally, the MSSM predicts
$M_{h^0} \lesssim 135$~GeV for the lighter of the two CP-even Higgses, in
remarkable agreement with Figure~\ref{stplot} (identifying $h^0$ with $H$).
The additional assumption of $R$-parity conservation (sufficient to forbid
proton decay by dimension~4 operators) implies that the lightest supersymmetric
particle is stable offering an explanation for the observed dark matter
(a similar statement applies to UED models).

\begin{theacknowledgments}
This work was supported by UNAM as DGAPA-PAPIIT project IN115207 and by CONACyT
(M\'exico) as project 82291-F.
\end{theacknowledgments}

\end{document}